\documentclass[newstyle,twocolumn,proceedings]{rmaa}

\usepackage{rmaacite}

\def\hii{H\thinspace{$\scriptstyle{\rm II}$}~}

\def\etal{{\it et al.}}

\def\eg{{\it e.g.},~}
\def\trit{$^3$H}
\def\3he{$^3$He}
\def\4he{$^4$He}
\def\6li{$^6$Li}
\def\7li{$^7$Li}
\def\bery{$^7$Be}
\def\Yp{Y$_{\rm P}$~}
\def\delY{$\Delta$Y~}
\def\omegab{$\Omega_{\rm B}$~}
\newcommand{\deln}{$\Delta N_\nu\;$}

\def\la{\mathrel{\mathpalette\fun <}}

\def\fun#1#2{\lower3.6pt\vbox{\baselineskip0pt\lineskip.9pt
  \ialign{$\mathsurround=0pt#1\hfil##\hfil$\crcr#2\crcr\sim\crcr}}}

\renewcommand{\P}[1]{%
\ifnum#1=1\hbox{OW~168--326E}\fi
\ifnum#1=2\hbox{OW~167--317}\fi
\ifnum#1=3\hbox{OW~163--317}\fi
\ifnum#1=5\hbox{OW~158--323}\fi
\ifnum#1=0\hbox{OW~171--334}\fi}

\title{CMB (And Other) Challenges To BBN}
\author{Gary Steigman\altaffilmark{1}, James P. Kneller, and 
Andrew Zentner
  \affil{Physics Department, The Ohio State University, Columbus, OH, USA} }
\altaffiltext{1}{Also, Astronomy Department, The Ohio State University.}

\fulladdresses{
\item Gary Steigman, James P. Kneller, and Andrew Zentner, Physics
Department, The Ohio State University, 174 West 18th Avenue, Columbus,
OH 43210, USA (steigman@mps.ohio-state.edu).}

\shortauthor{Steigman, Kneller, \& Zentner}
\shorttitle{Challenges To BBN}

\keywords{big bang nucleosynthesis --- primordial abundances --- baryon 
density --- cosmic microwave background anisotropies}

\abstract{%
  Primordial nucleosynthesis provides a probe of the universal 
  abundance of baryons when the universe was only a few minutes 
  old.  Recent observations of anisotropy in the cosmic microwave 
  background (CMB) probe the baryon abundance when the universe 
  was several hundred thousand years old.  Observations of type 
  Ia supernovae and clusters of galaxies in the very recent past, 
  when the universe is several billion years old and older, 
  provide a complementary measure of the baryon density in 
  excellent agreement with the early universe values.  The 
  general agreement among the three measurements represents 
  an impressive confirmation of the standard model of cosmology.  
  However, there is a hint that the CMB observations may not be 
  in perfect agreement with those from big bang nucleosynthesis 
  (BBN).  If this ``tension" between BBN and the CMB persists, 
  the standard model of cosmology may need to be modified.  Here, 
  in a contribution dedicated to Silvia Torres-Peimbert and Manuel 
  Peimbert, we describe how an asymmetry between neutrinos and 
  antineutrinos (``neutrino degeneracy") has the potential for 
  resolving this {\it possible} conflict between BBN and the CMB.}
  
\listofauthors{G.~Steigman, J.~P.~Kneller, \& A.~Zentner}
\indexauthor{Steigman, G.}
\indexauthor{Kneller, J.~P.}
\indexauthor{Zentner, A.}

\begin{document}

\maketitle

\section{Introduction}
\label{sec:intro}

Even though diamonds may not be forever, experimental constraints 
on proton stability are very strong ($\tau_{\rm N} > 10^{25}$ yr)
and baryon (nucleon) number should be preserved during virtually 
the entire evolution of the universe.  If so, then in the standard 
theories of particle physics and cosmology the baryon density 
at very early epochs is simply related to the baryon density 
throughout the later evolution of the universe.  In particular, 
the nucleon-to-photon ratio ($\eta \equiv n_{\rm N}/n_{\gamma}$) 
during primordial nucleosynthesis when the universe is only minutes 
old should be identical to $\eta$ measured when the universe 
is several hundred thousand years old and the cosmic microwave 
background (CMB) photons last scattered, as well as to $\eta$ 
in the present universe billions of years after the ``bang".  
Probing $\eta$ at such widely separated epochs in the evolution 
of the universe is a key test of the consistency of the standard 
models of particle physics and cosmology.  

The current status of this confrontation between theory and 
observations is reviewed here and our key results appear in 
Figure 1 where estimates of the universal baryon abundance 
at widely separated epochs are compared.  In \S~\ref{sec:bbn} 
the predicted BBN abundance of deuterium is compared with the 
primordial value inferred from observational data to derive 
the early-universe value of $\eta$.  After testing for the 
internal consistency of the standard model of Big Bang 
Nucleosynthesis (SBBN) by comparing the BBN-predicted and
observed abundances of the other light elements (\4he, \7li), 
an independent estimate of $\eta$ in the present (recent) 
universe is derived in \S~\ref{sec:sn1a} utilizing observations 
of clusters of galaxies and of type Ia supernovae (SNIa).  
These {\it independent} estimates of $\eta$ are compared to 
each other and, in \S~\ref{sec:cmb} to that from observations 
of the CMB anisotropy spectrum, a probe of $\eta$ in the 
several hundred thousand year old universe.  Having established 
that some ``tension" exists between $\eta_{\rm BBN}$ and 
$\eta_{\rm CMB}$, in \S~\ref{sec:nsbbn} a modification of 
SBBN involving ``degenerate" neutrinos is introduced and 
its consequences for the CMB anisotropies is explored.  
In \S~\ref{sec:disc} we summarize our conclusions.  The 
material presented here is extracted from our recent 
work \cite{swz,kssw} where further details and more 
extensive references may be found.

An alternate measure of the baryon abundance is the 
baryon density parameter, $\Omega_{\rm B}$, the ratio 
of the baryon mass density to the critical mass density.  
In terms of the present value of the Hubble parameter 
$h$ (H$_{0} \equiv 100h$~kms$^{-1}$Mpc$^{-1}$), and 
for a present CMB temperature of 2.725 K \cite{mather}, 
$\eta_{10} \equiv 10^{10}\eta = 274\Omega_{\rm B}h^{2}$.

\begin{figure}
  \includegraphics[width=\columnwidth]{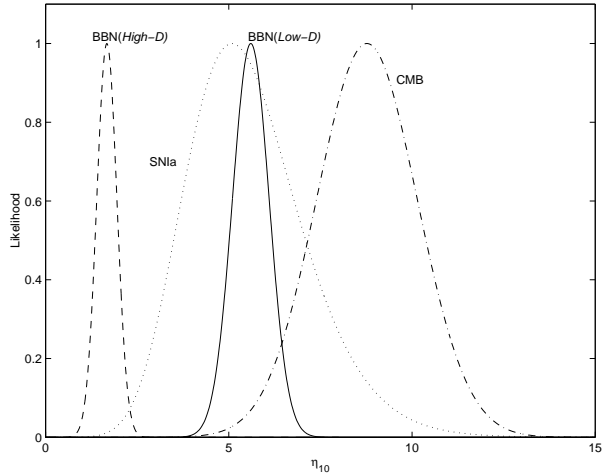}
  \caption{The likelihood distributions, normalized 
to unit maximum, for the baryon-to-photon ratio 
$\eta_{10} = 274\Omega_{\rm B}h^{2}$. The solid 
curve is the early universe value for low-D BBN 
while the dashed curve is for high-D BBN; the 
dotted curve (SNIa) is the present universe 
estimate; the dot-dashed curve shows the CMB 
inferred range.} 
  \label{fig:}
\end{figure}

\section{The Early Universe Baryon Density} 
\label{sec:bbn}

During its early evolution the universe is too hot to allow the
presence of astrophysically interesting abundances of bound nuclei
and primordial nucleosynthesis doesn't begin in earnest until the
temperature drops below $\approx 80$ keV, when the universe is a
few minutes old (for a recent review and further references see 
\pcite{osw}).  Prior to this time neutrons and protons have been
interconverting, at first rapidly, but more slowly after the first 
few seconds, driven by such ``charged-current" weak interactions as: 
$p + e^{-} \leftrightarrow n + \nu_{e}$, $n + e^{+} \leftrightarrow
p + \bar{\nu}_{e}$, and $n \leftrightarrow p + e^{-} + \bar{\nu}_{e}$ 
($\beta$-decay).  Once BBN begins neutrons and protons quickly 
combine to form deuterium which, in turn, is rapidly burned to 
\trit, \3he, and \4he.  There is a gap at mass-5 which, in the 
expanding, cooling universe, is difficult to bridge.  As a result, 
most neutrons available when BBN began are incorporated in the most 
tightly bound light nuclide, \4he.  For this reason, the \4he abundance 
(by mass, Y) is largely independent of the nuclear reaction rates 
but does depend on the neutron abundance at BBN which is determined 
by the competition between the weak interaction rates and the 
universal expansion rate (the early universe Hubble parameter, 
H).  In contrast, the abundances of D and \3he (\trit~is unstable, 
decaying to \3he) depend on the competition between 
the expansion rate and the nuclear reaction rates and, hence, on 
the baryon abundance $\eta$.  As a result, while D (and to a lesser 
extent, \3he) can provide a baryometer, \4he offers a test of the 
internal consistency of SBBN.  Although the gap at mass-5 is a barrier 
to the synthesis of heavier nuclides in the early universe, there is 
some production of mass-7 nuclei (\7li and \bery), albeit at a much 
suppressed level. The second mass gap at mass-8 eliminates (within 
SBBN) the synthesis of any astrophysically interesting abundances 
of heavier nuclides.  The abundance of lithium (after BBN, when 
the universe is sufficiently cool, \bery~will capture an electron 
and decay to \7li) is rate driven and can serve as a complementary 
baryometer to deuterium.

SBBN is overdetermined in the sense that for one adjustable parameter 
$\eta$, the abundances of four light nuclides (D, \3he, \4he, \7li) 
are predicted.  Here we concentrate on D and \4he.  Deuterium is an 
ideal baryometer candidate \cite{rafs} since it is only {\bf destroyed} 
(by processing in stars) in the post-BBN universe \cite{els}.  Deuterium 
is observed in absorption in the spectra of distant QSOs and its 
abundance in these high-redshift (relatively early in the star-forming 
history of the universe), low-metallicity (confirming that very little 
stellar processing has occurred) systems should represent the primordial 
value.  For three, high-z, low-Z QSO absorption-line systems a ``low" 
value of the deuterium abundance is found \cite{bt,omeara}, from which 
\scite{omeara} derive: D/H $= 3.0 \pm 0.4 \times 10^{-5}$.  Given the 
steep dependence of (D/H)$_{\rm BBN}$ on $\eta$ ($\propto \eta^{-1.6}$), 
this leads to a reasonably precise prediction for the baryon abundance 
at BBN: $\eta_{10} = 5.6 \pm 0.5$ ($\Omega_{\rm B}h^{2} = 0.020 \pm 
0.002$).  The likelihood distribution for this BBN-determined baryon
density is shown in Figure 1 by the curve labelled ``BBN(Low-D)".

\begin{figure}
  \includegraphics[width=\columnwidth]{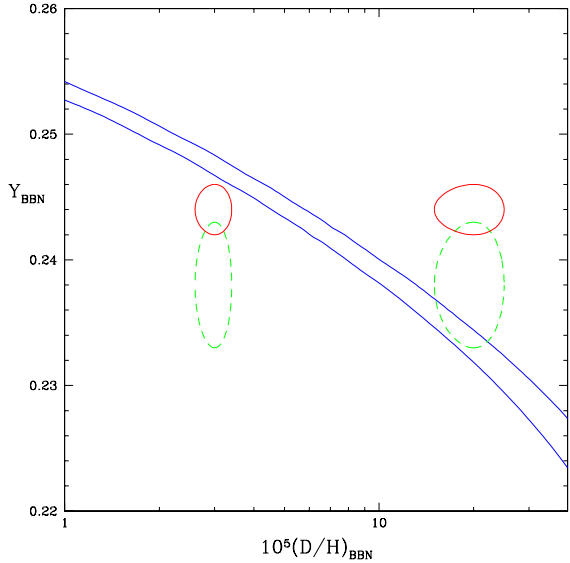}
   \caption{Comparison of the SBBN-predicted relation between the 
  primordial abundances of helium-4 (mass fraction, Y) and deuterium
  (ratio by number to hydrogen, D/H) and four sets of observationally
  inferred abundances.  The SBBN prediction, including uncertainties,
  is shown by the solid band.  The ``low-D" deuterium abundance is 
  from O'Meara \etal~(2000); the ``high-D" value is from Webb \etal 
  (1997).  The solid ellipses reflect the Izotov \& Thuan (1998) 
  helium abundance, while the dashed ellipses use the Olive, Steigman 
  \& Walker (2000) value.}
  \label{fig:bbn}
\end{figure}

Although any deuterium, observed anywhere (and at any time) in the
universe, provides a {\bf lower} limit to its primordial abundance, 
not all absorption identified with deuterium need actually be due 
to deuterium.  The absorption spectra of hydrogen and deuterium are 
identical, save for the wavelength/velocity shift (81 kms$^{-1}$) 
due to the very slightly different reduced masses.  Thus, any 
``observed" deuterium can only provide an {\bf upper} bound to 
the true deuterium abundance.  It is dismaying that such
crucial implications for cosmology rely at present on only three 
pieces of observational data.  Indeed, the most recently determined 
deuterium abundance \cite{omeara} is somewhat more than 3$\sigma$ 
lower than the previous primordial value based on the first two 
systems.  In fact, there is a fourth absorption-line system for 
which it has been claimed that deuterium is observed \cite{webb}.  
The deuterium abundance derived for this system is very high 
(``high-D"), nearly an order of magnitude larger than the low-D 
value, leading to a considerably smaller baryon abundance estimate.  
This determination suffers from a lack of sufficient data on the 
velocity structure of the absorbing cloud(s) and is a likely 
candidate for confusion with a hydrogen interloper masquerading 
as deuterium.  Nonetheless, for completeness, this estimate of 
the baryon density is included in Figure 1 by the curve labelled 
``BBN(High-D)".  We believe that the low-D value provides a better 
estimate of the true primordial abundance and, use it in the 
following for our ``preferred" estimate of the SBBN baryon density.
      
In Figure~\ref{fig:bbn} the band extending from upper left to lower 
right shows the relation between the SBBN-predicted abundances of D 
and \4he; the width of the band represents the (1$\sigma$) uncertainties 
in the predictions due to uncertainties in the nuclear and weak interaction 
rates.  Note that while D/H changes by an order of magnitude, Y hardly 
changes at all (\delY $\approx 0.015$).  Figure~\ref{fig:bbn} exposes 
the first observational challenge to SBBN.  For the observed (low-D) 
deuterium abundance (including its uncertainty), the SBBN-predicted 
helium abundance is Y $= 0.248 \pm 0.001$.  This is in disagreement 
with several determinations of the primordial helium abundance derived 
from observations of low-metallicity, extragalactic \hii regions.
From their survey of the literature \scite{os} find \Yp $= 0.234 
\pm 0.003$ (see also \pcite{oss} and \pcite{osw}), while from their 
own, independent data set \scite{it} derive \Yp $= 0.244 \pm 0.002$.  
It is clear that these results are in conflict and it is likely that 
unaccounted for systematic errors dominate the error budget.  For this 
reason a ``compromise" was advocated in \scite{osw}: \Yp $= 0.238 \pm 
0.005$.  Recently, in an attempt to either uncover or avoid some 
potential systematic errors, \scite{ppr} studied the nearby, albeit 
relatively metal-rich, \hii region NGC 346 in the SMC.  They found 
Y $= 0.2405 \pm 0.0018$ and, correcting for the evolution of Y with 
metallicity, derived \Yp $= 0.235 \pm 0.003$.  It is clear (see 
Fig.~\ref{fig:bbn}) that {\bf none} of these observational estimates 
is in agreement with the predictions of SBBN (low-D), although the 
gravity of the disagreement may be in the eye of the beholder.  The 
observationally inferred primordial helium abundance is ``too small" 
for the observationally determined deuterium abundance.  Either one 
(or both) of the abundance determinations is inaccurate at the level 
claimed, or some interesting physics (and/or cosmology) is missing 
from SBBN.  Notice that if the high-D abundance is the true primordial 
value there is no conflict between SBBN and the \scite{osw} helium 
abundance, while the \scite{it} abundance is now too high.  Before 
addressing the role of possible non-standard BBN in relieving the 
tension between D and \4he, other, non-BBN, bounds on the baryon 
abundance are considered and compared to $\Omega_{\rm BBN}$. 

\section{The Present Universe Baryon Density}
\label{sec:sn1a}

It is notoriously difficult to inventory baryons in the present universe.  
\scite{ps} have attempted to find the density of those baryons which 
reveal themselves by shining (or absorbing!) in some observationally 
accessible part of the electromagnetic spectrum: ``luminous baryons".  
It is clear from \scite{ps} that most baryons in the present universe 
are ``dark" since they find $\Omega_{\rm LUM} \approx 0.0022 + 
0.0006h^{-1.3} \ll \Omega_{\rm BBN}$.  At the very least this lower 
bound to $\Omega_{\rm B}$ is good news for SBBN, demonstrating that 
the baryons present during BBN {\it may} still be here today.  In a 
more recent inventory which includes some estimates of dark baryons, 
\scite{fhp} find a larger range ($0.007 \la \Omega_{\rm B} \la 0.041$) 
that has considerable overlap with $\Omega_{\rm BBN}$.  

A complementary approach to the present universe baryon density is 
to combine an estimate of the {\it total} mass density, baryonic 
plus ``cold dark matter" (CDM), $\Omega_{\rm M}$, with an independent 
estimate of the universal baryon {\it fraction} $f_{\rm B}$ to 
find $\Omega_{\rm B} = f_{\rm B}\Omega_{\rm M}$.  Recently, we 
\cite{swz} imposed the assumption of a ``flat" universe and used 
the SNIa magnitude-redshift data \cite{sn1a} to find $\Omega_{\rm M}$ 
($0.28^{+0.08}_{-0.07}$), which was combined with a baryon fraction 
estimate ($f_{\rm B}h^{2} = 0.065^{+0.016}_{-0.015}$) based on X-ray 
observations of rich clusters of galaxies \cite{xray} and the HST Key 
Project determination of the Hubble parameter ($h = 0.71 \pm 0.06$; 
\pcite{hst}) to derive $\eta_{10} = 4.8^{+1.9}_{-1.5}$ ($\Omega_{\rm 
B}h^{2} = 0.018^{+0.007}_{-0.005}$).  Subsequently \scite{grego}, 
utilizing observations of the Sunyaev-Zeldovich effect in X-ray 
clusters, have reported a very similar value for the cluster hot 
gas fraction to that adopted in \scite{swz}.  For the \scite{grego} 
value for $f_{\rm B}$, which may be less vulnerable to systematics, 
the present universe baryon density is, $\eta_{10} = 5.1^{+1.8}_{-1.4}$ 
($\Omega_{\rm B}h^{2} = 0.019^{+0.007}_{-0.005}$).  This distribution 
is shown in Figure 1 by the curve labelled SNIa.  Although the 
uncertainties in this estimate at $z \approx 0$ are large, the 
excellent overlap lends support to the low-D SBBN baryon abundance.  
The poor overlap with the high-D SBBN baryon density argues against the 
high D/H being representative of the primordial deuterium abundance.

\section{The Baryon Density At $z \sim 1000$}
\label{sec:cmb}

At redshift $z \sim 1000$, when the universe is several hundred
thousand years old, the temperature of the CMB radiation has
cooled sufficiently for neutral hydrogen (and helium) to form.  
The CMB photons are now freed from the tyranny of electron 
scattering and they propagate freely carrying the imprint of 
cosmic perturbations as well as encoding the parameters of the 
cosmological model, in particular the baryon density.  Observations 
of the CMB anisotropies therefore provide a probe of \omegab at 
a time in the evolution of the universe intermediate between BBN 
and the present epoch.

Recent observations of the CMB fluctuations by the BOOMERANG 
\cite{boom} and MAXIMA \cite{max} experiments have provided 
a means for constraining the baryon density at $z \sim 1000$.  
The relative height of the first two ``acoustic peaks" in 
the CMB anisotropy spectrum is sensitive to the baryon density.  
Although the precise value of $\Omega_B h^2$ depends on the 
choice of ``priors" for the other cosmological parameters 
which must be included in the analysis, the CMB-inferred
baryon density exceeds that derived from BBN (low-D) by 
$\sim 50\%$, $\Omega_B h^2 \sim 0.03$ ($\eta_{10} \sim 8$).  
The baryon density likelihood distribution shown in Figure 
1 is based on the combined Boomerang and Maxima analysis 
of \scite{jaffe} who find $\Omega_B h^2 = 0.032 \pm 0.005$ 
($\eta_{10} = 8.8 \pm 1.4$). 

It is clear from Figure 1 that while there is excellent overlap
between the low-D SBBN and SNIa baryon density estimates, the 
high-D SBBN value is discordant.  Furthermore, there is a hint 
that the CMB value may be too large.  Note that the apparent
``agreement" (or, minimal apparent disagreement) in Figure 1 
is an artifact of normalizing each likelihood function to unit 
maximum.  In fact, the CMB data excludes the central value of
low-D SBBN at greater than 98\% confidence.
Although it may well be premature to take this ``threat" to SBBN 
seriously, this potential discrepancy has led to the suggestion 
that new physics may need to be invoked to reconcile the BBN and 
CMB predictions for $\Omega_B h^2$.  This possibility is discussed 
next.

\section{Beyond SBBN}
\label{sec:nsbbn}

\begin{figure}
  \includegraphics[width=\columnwidth]{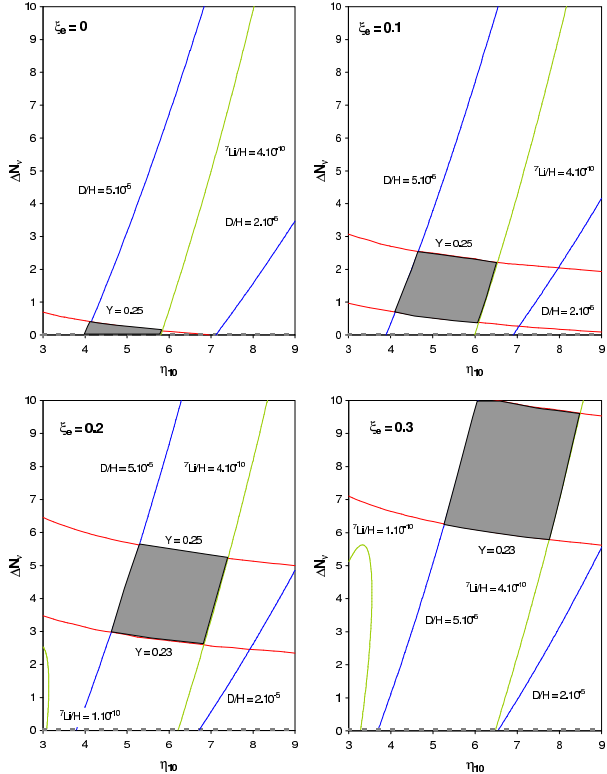}
  \caption{Iso-abundance contours for deuterium 
(D/H), lithium (Li/H) and helium (mass fraction, Y) in the 
\deln -- $\eta_{10}$ plane for four choices of $\nu_{e}$ 
degeneracy ($\xi_{e}$).  The shaded areas highlight the range
of parameters consistent with the adopted abundance ranges.}
  \label{fig:iso}
\end{figure}

Observations of deuterium and helium (and, perhaps lithium) offer 
the first challenge to SBBN and the baryon density derived from it 
(see \S~\ref{sec:bbn}).  Setting aside the very real possibility 
of errors in the observationally derived abundances, how might 
SBBN be modified to account for a helium abundance which is
(predicted to be) too large?  Not surprisingly, the options are 
manifold.  One possibility is to modify the expansion rate of the 
early universe.  If for some reason the universe were to expand 
more slowly than in the standard model, there would be more time 
for neutrons to convert to protons, resulting in a lower primordial 
helium abundance.  In addition, a slower expansion would leave more 
time for deuterium to be destroyed resulting in a lower D-abundance.  
To compensate for this, the BBN baryon density would need to be 
{\bf reduced}.  This has the further beneficial effect of reducing 
the predicted lithium abundance, as well as reducing (very slightly) 
the predicted helium abundance.  Thus, a {\bf slower} expansion 
rate in the early universe can reconcile the predicted and observed 
deuterium and helium abundances (c.f. \pcite{chen,zia}).  But, since 
this ``solution" requires a {\bf lower} baryon density, it exacerbates 
the tension between BBN and the CMB.

Although a {\bf speed up} in the expansion rate offers the 
possibility of reconciling the observed deuterium abundance with 
the high baryon density favored by the CMB, it greatly exacerbates 
the helium abundance discrepancy and increases the tension between 
the predicted and observed lithium abundances.  To reconcile the 
BBN and CMB estimates of the baryon density, while maintaining 
(or, establishing!) consistency between the predicted and observed 
primordial abundances, additional ``new physics" needs to be invoked.

The simplest possibility for reducing the BBN-predicted helium 
abundance is a non-zero chemical potential for the electron 
neutrinos.  An excess of $\nu_{e}$ over $\bar{\nu}_{e}$ can 
drive down the neutron-proton ratio, leading to reduced 
production of helium-4.  Thus, one path to reconciling 
BBN with a high baryon density is to ``arrange" for a 
faster than standard expansion rate ($S \equiv H'/H > 
1$) {\bf and} for degenerate $\nu_{e}$.  Although these 
two effects need not be related, neutrino degeneracy can, 
in fact, provide an economic mechanism for both since the 
energy density contributed by degenerate neutrinos exceeds 
that from non-degenerate neutrinos, leading to an enhanced 
expansion rate during radiation-dominated epochs ($H'/H = 
(\rho '/\rho)^{1/2} > 1$).  Thus, one approach to non-standard 
BBN is to introduce two new parameters, the speed up factor 
${\bf S}$ and the electron neutrino degeneracy parameter 
${\bf \xi_e}$, where $\xi_e = \mu_e/T_\nu$ is the ratio of 
the electron neutrino chemical potential $\mu_e$ to the neutrino 
temperature $T_\nu$.  For degenerate neutrinos the energy density 
($\rho_{\nu}(\xi)$) exceeds the non-degenerate energy density 
($\rho_{\nu}^{0}$) 

\begin{equation}
\Delta\rho_{\nu}/\rho_{\nu}^{0} = 15/7[(\xi/\pi)^4 + 2(\xi/\pi)^2].
\end{equation}
Thus, neutrino degeneracy has the same effect (on the early universe
expansion rate) as would additional species of light, non-degenerate
neutrinos.  In terms of the equivalent number of ``extra",  
non-degenerate, two-component neutrinos {\bf $\Delta N_\nu$}, 
the speed up factor is

\begin{equation}
S = (1 + 7\Delta N_\nu/43)^{1/2}.
\end{equation}
To facilitate comparison with the published literature, \deln is 
used in place of $S$.  Since $\Delta N_\nu = \Delta\rho_{\nu}/
\rho_{\nu}^{0}$, \deln accounts for the additional energy density 
contributed by all the degenerate neutrinos (see eq.~1) {\it as 
well as any other energy density not accounted for in the standard 
model of particle physics} (e.g., additional relativistic particles) 
expressed in terms of the equivalent number of extra, non-degenerate, 
two-component neutrinos.  However, our results are independent 
of whether \deln (or the corresponding value of $S$) arises from 
neutrino degeneracy, from ``new''  particles, or from some other 
source.  Note that a non-zero value of $\xi_e$ implies a non-zero 
contribution to $\Delta N_\nu$ from the electron neutrinos alone.  
This contribution has been included in our calculations.  However, 
for the range of $\xi_e$ which proves to be of interest 
($\xi_e \la 0.5$; see Fig.~3), the degenerate electron 
neutrinos contribute only a small fraction of an additional neutrino 
species to the energy density ($\Delta N_\nu \la 0.1$).  As \scite{ks} 
and \scite{ostw} have shown, the observed primordial abundances of 
the light nuclides can be reconciled with very large baryon densities 
provided that $\xi_{e} > 0$ and \deln is sufficiently large. 

The parameter space \scite{kssw} investigated is three-dimensional: 
$\eta$, $\xi_{e}$, and $\Delta N_\nu$.  Generous ranges for the 
primordial abundances were chosen which are large enough to encompass 
systematic errors in the observations, as well as to account for 
the BBN uncertainties due to imprecisely known nuclear and/or 
weak reaction rates: $0.23 \le {\rm Y}_P \le 0.25$, $2 \times 
10^{-5} \le {\rm D/H} \le 5 \times 10^{-5}$, $1 \times 10^{-10} 
\le {\rm ^7Li/H} \le 4 \times 10^{-10}$.  Since we wish to compare 
to the predictions of the CMB, which are sensitive to $\eta$ and 
$\Delta N_\nu$, but independent of $\xi_e$, the allowed BBN region 
is projected onto the $\eta - \Delta N_\nu$ plane.  The BBN results 
are shown in the four panels of Figure 3 where, for four choices 
of $\xi_e$ the iso-abundance contours for Y$_{P}$, D/H and Li/H 
are shown.  The shaded areas 
highlight the acceptable regions in our parameter space.  As 
$\xi_e$ increases, the allowed region moves to higher values of 
$\eta$ and $\Delta N_\nu$, tracing out a BBN-consistent band in 
the $\eta - \Delta N_\nu$ plane.  This band is shown in Figure 
4 where the CMB constraints on the same parameters  (under the 
assumption of a flat universe; for details and other cases, see 
\pcite{kssw}) are shown.  The trends are easy to understand: as 
the baryon density increases the universal expansion rate (measured 
by $\Delta N_\nu$) increases to keep the deuterium and lithium 
unchanged, while the $\nu_{e}$ degeneracy ($\xi_{e}$) increases 
to maintain the helium abundance at its (correct!) BBN value. 

\begin{figure}
 \includegraphics[width=\columnwidth]{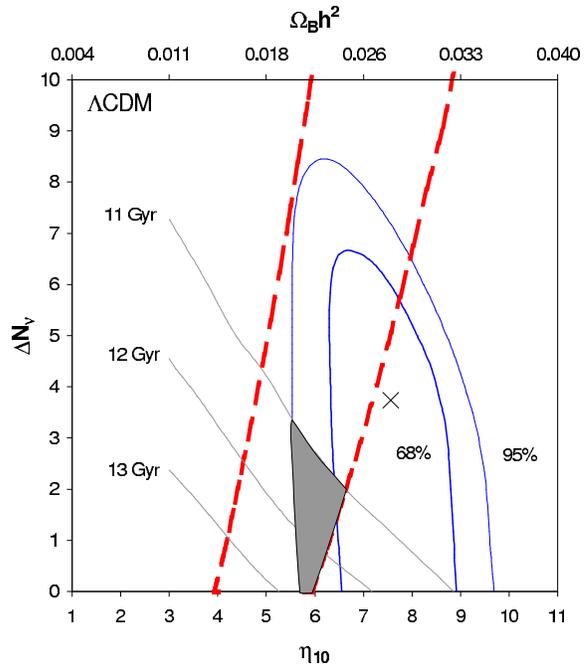}
  \caption{The BBN (dashed) and CMB (solid) 
	contours (flat, $\Lambda$CDM model) in 
	the \deln -- $\eta_{10}$ plane.  The 
	corresponding best fit iso-age contours 
	are shown for 11, 12, and 13 Gyr.  The 
	shaded region delineates the parameters 
	consistent with BBN, CMB, and t$_{0} > 
	11$ Gyr.}
  \label{fig:comp}
\end{figure}

The CMB anisotropy spectrum depends on the baryon density and
on the universal expansion rate (through the relativistic
energy density as measured by $\Delta N_{\nu}$) as well as 
on many other cosmological parameters which play no role in 
BBN.  But, in fitting the CMB data, choices must be made 
(``priors") of the values or ranges of these other parameters.  
In \scite{kssw} several cosmological models and several choices 
for the ``priors" were explored.  Figure 4 shows the BBN/CMB 
comparison for the ``flat, $\Lambda$CDM" model (Case C of 
\pcite{kssw}).  The significant overlap between the BBN-allowed 
band and the CMB contours, confirms that if we allow for ``new 
physics" ($\xi_{e} > 0$ and \deln $> 0$), the tension between 
BBN and the CMB can be relieved.  

Since the points in the $\eta$ -- \deln plane are projections 
from a multi-dimensional parameter space, the relevant values 
of the ``hidden" parameters may not always be consistent with 
other, independent observational data which could provide 
additional constraints.  As an illustration, three iso-age 
contours (11, 12, and 13 Gyr), are shown in Figure 4.  The 
iso-age trend is easy to understand since as \deln increases, 
so too do the corresponding values of the matter density 
($\Omega_{M}$) and the Hubble parameter (H$_{0}$) which minimize 
$\chi^{2}$.  Furthermore, since $\Omega_{M} + \Omega_{\Lambda} = 
1$, $\Omega_{\Lambda}$ decreases.  All of these lead to younger 
ages for larger values of $\Delta N_\nu$.  Note that if an age 
constraint is imposed (\eg that the universe today is {\bf at 
least} 11 Gyr old \cite{chaboyer}), then the BBN and CMB overlap 
is considerably restricted (to the shaded region in Figure 4).  
Even with this constraint it is clear that for modest ``new 
physics'' (\deln $\la 4$; $\xi_{e} \la 0.3$) there is a small 
range of baryon density ($0.020 \la \Omega_{B}h^{2} \la 0.026$) 
which is concordant with both the BBN and CMB constraints, as 
well as the present universe baryon density.

\section{Summary and Conclusions}
\label{sec:disc}

According to the standard models of cosmology and particle physics, 
as the universe evolves from the first few minutes to the present, 
the ratio of baryons (nucleons) to photons, $\eta$, should be 
unchanged.  The abundance of deuterium, a relic from the earliest 
epochs, identifies a nucleon abundance $\eta_{10} \sim 5.6$.  The 
CMB photons, relics from a later, but still distant epoch in the
evolution of the universe suggest a somewhat higher value, $\eta_{10} 
\sim 8.8$.  Although most baryons in the present universe are dark 
and the path to the current nucleon-to-photon ratio is indirect, our 
estimates suggest $\eta_{10} \sim 5.1$.  That these determinations 
are all so close to one another is a resounding success of the 
standard model.  The possible differences may either reflect the 
growing pains of a maturing field whose predictions and observations 
are increasingly precise, or perhaps, be pointing the way to new 
physics.  Exciting times indeed!

\acknowledgements We are pleased to acknowledge Bob Scherrer and 
Terry Walker for their contributions to the work reviewed here.  
Financial support for this research at OSU has been provided by 
the DOE (DE-FG02-91ER-40690).


\end{document}